\documentclass[aps,prl,floatfix,twocolumn,showpacs,reprint,superscriptaddress]{revtex4-1}
\usepackage{amsfonts}
\usepackage{mathrsfs}
\usepackage{amsmath}
\usepackage{color}
\usepackage{graphicx}
\usepackage{bm}
\usepackage{amssymb}
\usepackage{xspace}
\usepackage{epstopdf}
\usepackage{dcolumn}
\usepackage{longtable}
\usepackage{multirow}
\usepackage[colorlinks=true, letterpaper=true, pdfstartview=FitV, linkcolor=blue, citecolor=blue, urlcolor=blue]{hyperref}

\makeatletter

\newcommand{\Rmnum}[1]{\expandafter\@slowromancap\romannumeral #1@}
\makeatother
\begin{document}
\title{Dirac and Weyl Superconductors in Three Dimensions}

\author{Shengyuan A. Yang}
\affiliation{Engineering Product Development, Singapore University of Technology and Design, Singapore 138682, Singapore}

\author{Hui Pan}
\affiliation{Department of Physics, Beihang University, Beijing 100191, China}

\author{Fan Zhang}
\email{zhf@sas.upenn.edu}
\affiliation{Department of Physics and Astronomy, University of Pennsylvania, Philadelphia, Pennsylvania 19104, USA}
\affiliation{Kavli Institute for Theoretical Physics, University of California, Santa Barbara, California 93106, USA}

\begin{abstract}
We introduce the concept of 3D Dirac (Weyl) superconductors (SC),
which have protected bulk four(two)-fold nodal points and surface Andreev arcs at zero energy.
We provide a sufficient criterion for realizing them in centrosymmetric SCs with odd-parity pairing and mirror symmetry,
e.g., the nodal phases of Cu$_x$Bi$_2$Se$_3$.
Pairs of Dirac nodes appear in a mirror-invariant plane when the mirror winding number is nontrivial.
Breaking mirror symmetry may gap Dirac nodes producing a topological SC.
Each Dirac node evolves to a nodal ring when inversion-gauge symmetry is broken.
A Dirac node may split into a pair of Weyl nodes, only when time-reversal symmetry is broken.
\end{abstract}
\maketitle

Topological states of matter have attracted significant attention since the discovery of topological insulators~\cite{RMP1}.
The idea of topological classification was soon generalized to superconductors (SC) which have energy gaps for quasiparticles~\cite{RMP2}.
Interestingly, topological phases also exist for systems without energy gaps.
Graphene and its ABC stacked cousins are examples of two dimensional (2D) semimetals~\cite{ABC1,ABC2},
in which their Fermi surfaces consist of isolated points that are protected by the chiral (sublattice) symmetries.
Indeed, the constant energy surfaces of these graphene few-layers have winding numbers set by the number of layers.
Recently, the topological semimetal concept has been extended to three dimensions (3D).
Unlike the critical point between 3D trivial and topological insulators with inversion symmetry,
the fourfold degenerate Fermi points in 3D Dirac semimetals~\cite{Rappe1,Fang1,Fang2,Rappe2,ex1,ex2,ex3} are protected by crystalline symmetries.
When an essential symmetry is broken, a Dirac semimetal, both in 2D and 3D, may become a topological or a trivial insulator.
Moreover, a Dirac point may split into two Weyl points when inversion or time-reversal symmetry (TRS) is broken,
and the Dirac semimetal becomes a Weyl semimetal.
In 3D a pair of Weyl points~\cite{w1,w2,w3,w4,w5} is protected by Chern numbers $\pm1$ of constant energy surfaces with each enclosing one Weyl node,
leading to a surface Fermi arc.
In contrast, a 3D Dirac semimetal may or may not have a surface Fermi arc.
However, it is generally not an easy task to pin the Fermi energy exactly at the nodal points in semimetals.

Nodal phases are common for unconventional SCs. One may naturally wonder whether there also exist Dirac or Weyl SCs in 3D.
If the answer is positive, what protect their nodes and are there any surface consequences?
Remarkably enough, we discover that their existence is indeed possible. In this Letter,
we will provide a sufficient criterion for their realizations and will discuss their topological protections and surface consequences.

In the Bogliubov-de Gennes (BdG) description of SCs,
the particle-hole redundancy leads to a natrual half-filling and an intrinsic particle-hole symmetry (PHS).
Compared to the cases of semimetals, the former feature simplifies our task to focus on the nodal points at zero energy,
whereas the latter feature poses an additional symmetry constraint which plays intriguing roles in stabilizing the nodes.
Specifically, we find that a 3D Dirac SC can be realized in a nodal phase of a centrosymmetric SC with odd-parity pairing and mirror symmetry.
Pairs of Dirac nodes would appear in a mirror-invariant plane when the mirror winding number~\cite{zhang2013m,yao2013} is nontrivial.
Each Dirac node is protected locally by the combination of mirror symmetry, TRS, and an inversion-gauge symmetry
which we will introduce in a short while.

Breaking any symmetry destroys the Dirac nodes:
(i) breaking mirror symmetry may fully gap the nodes producing a topological SC;
(ii) breaking inversion-gauge symmetry extends each Dirac node to a robust nodal ring;
(iii) a Dirac node may split into a pair of Weyl nodes only when TRS is broken.
These evolutions of bulk Dirac nodes and the corresponding deformations of surface Andreev arcs are summarized in Fig.~\ref{fig2}.
Our physics might be realized in the nodal phase of Cu$_x$Bi$_2$Se$_3$~\cite{fu2010},
which at least serves as a concrete example to illustrate the new physics we will present.

\begin{figure*}
\scalebox{0.28}{\includegraphics*{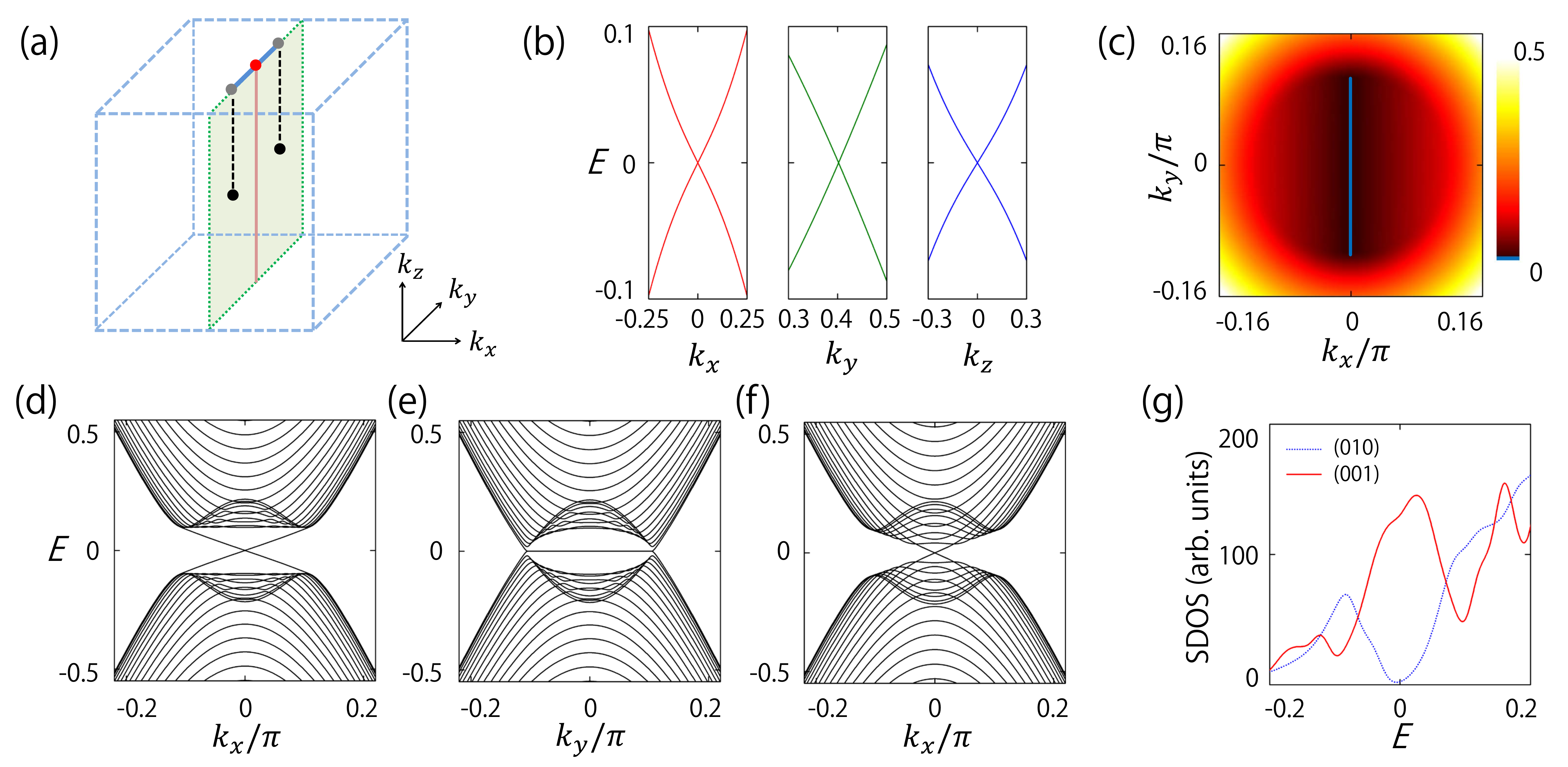}}
\caption{Features of a 3D Dirac superconductor. (a) Two Dirac nodes (black dots) located in the (shaded) mirror invariant plane. The nontrivial mirror winding number of the $k_x,k_y=0$ (pink) line protects the nodes and dictates the presence of a Majorana Kramers pair (red dot) and Majorana arc (blue curve) on the $(001)$ surface. (b) The linear energy dispersion around one Dirac node from our tight binding calculations~\cite{parameter-fig1}; each band is doubly degenerate because of TRS and inversion-gauge symmetry. (c), (d), and (e) Band structures of a $(001)$ slab. (c) Positive energy dispersion of the surface states. The blue color corresponds to zero energy. (f) Band structures of a $(010)$ slab. (g) Surface density of states for $(001)$ and $(010)$ surfaces.}
\label{fig1}
\end{figure*}

Our analysis starts from the Fu-Berg mean-field model~\cite{fu2010} describing the SC state of Cu$_x$Bi$_2$Se$_3$:
\begin{equation}\label{BdG}
H=\left[v({\bm k}\times{\bm s})\cdot{\hat z}\,\sigma_z+v_z k_z \sigma_y+m\sigma_x-\mu\right]\tau_z+{\bm\Delta}\tau_x\,.
\end{equation}
In the above BdG Hamiltonian the Pauli matrices $\bm\sigma$, $\bm s$, and $\bm\tau$ act on the orbital, spin, and Nambu spaces, respectively.
For each orbital we have chosen the basis as $(c_{\bm k\uparrow},c_{\bm k\downarrow},c^\dagger_{-\bm k\downarrow},-c^\dagger_{-\bm k\uparrow})^T$.
The $\tau_z$-term describes the normal state near $\Gamma$ point, with its form determined~\cite{zhan2012,liu2010} by the inversion ($\mathcal{P}=\sigma_x$) symmetry, the mirror ($\mathcal{M}=-is_x$) symmetry, i.e.,
\begin{eqnarray}
\mathcal{M}H(k_x,k_y,k_z)\mathcal{M}^{-1}=H(-k_x,k_y,k_z)\,,
\end{eqnarray}
and the $\mathcal{C}_3(\hat{z})$ symmetry of the Cu$_x$Bi$_2$Se$_3$ crystal. $\hat{z}$ is the quintuple-layer normal,
$m$ is the normal state band gap and $\mu$ is the chemical potential.
The pairing term $\bm\Delta$ can be classified~\cite{fu2010} according to the representations of the crystal point group $D_{3d}$.
The existence of mirror symmetry requires~\cite{zhang2013m} the winding number to vanish for any fully gapped 3D SC with TRS.
Indeed Fu and Berg have shown~\cite{fu2010} that the topological state (${\bm\Delta}\sim\sigma_y s_z$) breaks the mirror symmetry
whereas the states respecting the symmetry are either trivial (${\bm\Delta}\sim I$ or $\sigma_x$) or nodal.
Our focus will be on the nodal phase with ${\bm\Delta}\sim\sigma_y s_{x}$. Similar results also apply to other nodal phases.

Under inversion the normal state has even parity whereas the pairing $\sigma_y s_x\tau_x$ has odd parity,
yet the Hamiltonian~(\ref{BdG}) still has an inversion-gauge symmetry, i.e.,
\begin{eqnarray}
\tau_z\mathcal{P}H(\bm k)\mathcal{P}^{-1}\tau_z=H(-\bm k)\,.
\end{eqnarray}
We observe that in the mirror invariant plane with $k_x=0$, the nodal points are located away from the time-reversal and mirror invariant line $k_y=0$. Indeed, along this special line, the BdG Hamiltonian~(\ref{BdG}) reduces to
\begin{equation}
H=(v_z k_z\sigma_y+m\sigma_x-\mu)\tau_z+\Delta \sigma_y s_x\tau_x\,,
\label{h4}
\end{equation}
the spectrum of which is fully gapped as long as the pairing potential $\Delta$ is nonzero. It is easy to verify that Eq.~(\ref{h4}) is adiabatically connected to the case for $m=\mu=0$ (by first letting $m\rightarrow 0$ then $\mu \rightarrow0$) described by
\begin{equation}\label{h4a}
\bar{H}=v_z k_z\sigma_y\tau_z+\Delta \sigma_y s_x\tau_x\,,
\end{equation}
in which both $\sigma_y$ and $s_x$ are good quantum numbers with eigenvalues $\pm 1$. Now consider the interface between a 1D system described by Eq.~(\ref{h4a}) and the vacuum. The trivial vacuum is adiabatically connected to a pure s-wave SC and thus can be modeled by $\Delta_s \tau_x$, with $\Delta_s\rightarrow\infty$ and $\Delta_s\cdot\Delta>0$~\cite{zhan2012,zhang2013t}. One recognizes that out of the four possible cases $\sigma_y=\pm 1$, $s_x=\pm 1$, there are two copies of Jackiw-Rebbi problem~\cite{jack1976} of a 1D two-band Dirac model, i.e., one with $\sigma_y=-1$ and $s_x=1$ and the other with $\sigma_y=1$ and $s_x=-1$. Each Jackiw-Rebbi problem has a zero-energy bound state localized on the surface~\cite{jack1976,zhan2012}. Due to the TRS and PHS, these two zero modes form a Majorana Kramers pair at $k_x=k_y=0$ at the top or bottom surface Brillouin zone (BZ), as sketched in Fig.~\ref{fig1}(a). Moreover, across this special momentum, there exists a Majorana Kramers arc connecting the projected nodal points in the surface state spectrum.

To visualize the Dirac nodes and their surface consequences more clearly, we numerically calculate the energy spectrum for a slab geometry terminated by vacuum at proper surfaces, using a layered hexagonal lattice model~\cite{lattice}.  Fig.~\ref{fig1}(c-e) show the features of $(001)$ surface states. The two surface projected nodal points are located in the $k_x=0$ line with finite $k_y$ values in the surface BZ; there is a flat surface arc connecting the projected nodal points. This arc is doubly degenerate and hosts a Majorana Kramers pair at its center.

Remarkably, the presence and the flatness of Majorana Kramers arc in Fig.~\ref{fig1} are not accidental and indeed protected by the mirror symmetry and TRS, as we explain now. Consider the mirror invariant plane ($k_x=0$) with a pair of nodal points at $k_y=\pm k_n$.
In this plane, two mirror subspaces ($s_x=\pm$) decouple and are related by TRS and PHS. Both TRS and PHS are broken in each mirror subspace, yet the chiral symmetry, i.e. the product of TRS and PHS, is still respected. For any fully gapped loop $l$ in this plane, the presence of chiral symmetry allows the definitions of total and mirror winding numbers~\cite{zhang2013m,gauge} as follows:
\begin{equation}
\gamma_{t,m}=\frac{1}{2\pi}\oint_{l}(\mathcal{A}_{{\bm k}+}\pm\mathcal{A}_{{\bm k}-})\cdot \text d{\bm k}\,,
\label{mbp}
\end{equation}
where $\mathcal{A}_{{\bm k}\pm}$ is the Berry connection of the negative energy band in the $s_x=\pm$ mirror subspace.
Explicit calculations show that $\gamma_m=1$~\cite{regular} and $\gamma_t=0$ for Eq.~(\ref{h4}),
which dictates the presence of a Majorana Kramers pair~\cite{zhang2013t} at the surface BZ center.
In the $k_x=0$ plane, the states in $k_y=a$ loops with $|a|<k_n$ and with $|a|>k_n$ are adiabatically connected to the state of Eq.~(\ref{h4}) and to the vacuum state~\cite{pi}, respectively. Hence in the former case $\gamma_m=1$ whereas in the latter case $\gamma_m=0$. Equivalently, $\gamma_m=1$ ($-1$) for any loop enclosing the $k_y=k_n$ ($-k_n$) nodal point whereas $\gamma_m=0$ for any loop enclosing both or neither nodal points. The nontrivial $\gamma_m$ has two important consequences.
\begin{figure}[t!]
\scalebox{0.26}{\includegraphics*{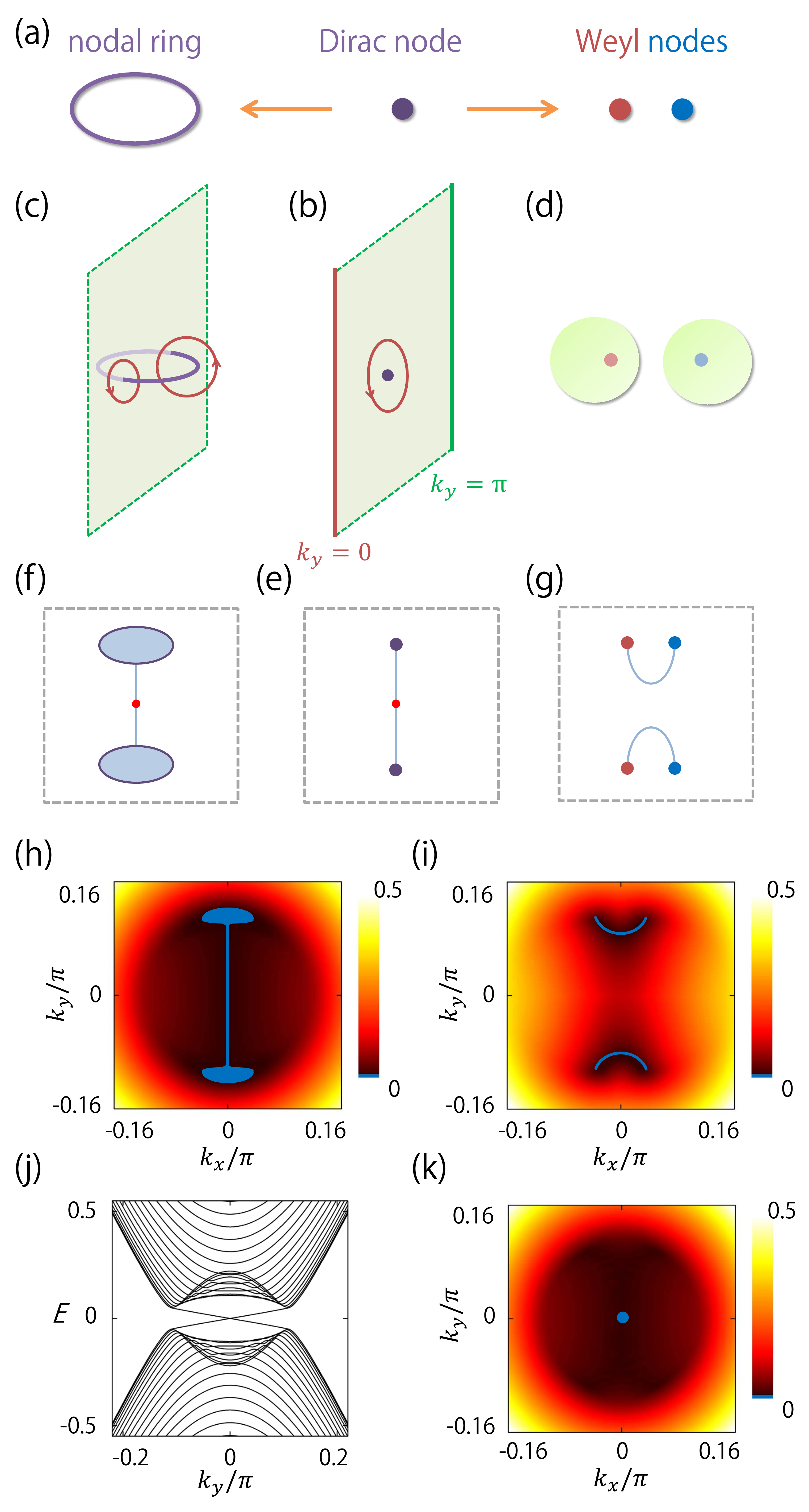}}
\caption{Consequences of symmetry breaking for a Dirac SC. (a) When the inversion-gauge symmetry or TRS is broken, a Dirac node evolves into a nodal ring or two Weyl nodes normal to the mirror invariant plane, respectively. (b-d) A Dirac node (nodal ring) protected by the mirror (total) winding number $\pm 1$ of the surrounding loop (red); a Weyl node protected by the Chern number $\pm 1$ of the surrounding sphere (light green).(e-g) Sketches of the surface zero-energy modes for each case. (e) Majorana Kramers arc (light blue) connecting the two surface projected Dirac nodes. (f) Two surface projected nodal rings with an extended area of zero energy states inside. (g) A surface arc connecting a pair of surface projected Weyl nodes. (h-k) Tight binding calculations~\cite{parameter-fig1} for a $(001)$ slab. (h) and (i) correspond to the scenarios in (f) and (g), respectively. (j) and (k) show that a Dirac SC may become a fully gapped topological SC when the mirror symmetry is broken. The blue color corresponds to the zero energy. We have chosen $\sigma_z\tau_z$ to break the inversion-gauge symmetry, $\sigma_x s_x$ to break TRS, and $\sigma_ys_z\tau_x$ to break the mirror symmetry, respectively.}
\label{fig2}
\end{figure}

First, the nodal points must exist in the mirror invariant plane. To demonstrate this, suppose that there is no nodal point in the plane, then the presence of chiral symmetry requires that both the total and the mirror Berry curvatures~\cite{zhang2013m} must vanish, leading to a contradicting result $\gamma_m(0)=\gamma_m(\pi)$ via Stokes' theorem. Therefore, the derived difference in $\gamma_m$ implies the presence of a pair of nodal points in the plane. Out of the plane, the absence of mirror invariance implies gap opening. Fig.~\ref{fig1}(b) shows that the quasiparticle energy dispersion is linear in all directions near the bulk nodal point. Furthermore, each band must be fourfold degenerate because of the presence of both TRS and inversion-gauge symmetry. Therefore, the nodes are Dirac nodes.

Secondly, $\gamma_m=1$ amounts to a Berry phase $\pm\pi$ in the decoupled mirror subspace with $s_x=\pm$. This implies a protected surface state at $(k_x=0,k_y)$ for any $|k_y|<\pm k_n$ in any surface that preserves the mirror symmetry. Moreover, the presence of chiral symmetry in each mirror subspace pins the surface state to zero energy. Therefore, the Majorana arc must be dispersionless and spin degenerate.

From the above discussion, evidently, the combination of TRS, mirror symmetry,
and inversion-gauge symmetry provides the protection of the Dirac nodes.
When either of these symmetries is broken, the Dirac nodes become unstable.
To facilitate the understanding of the consequence of symmetry breaking,
we construct a local effective model near a Dirac node from general symmetry analysis.
Since a Dirac node lies in a mirror invariant plane, the effective model must have the following {\em local} symmetries which commute with each other: chiral symmetry $\Pi$, mirror symmetry $\mathcal{M}$, and the product of TRS and inversion-gauge symmetry $\mathcal{W}$.
For a Dirac node, the low-energy subspace has a dimension of four. We choose the representations of symmetry operations as follows:
$\Pi=\tau_y$, $\mathcal{M}=-is_x$, and $\mathcal{W}=s_y\tau_z K$ with $K$ the complex conjugation.
Note that the Pauli matrices $\tau$ and $s$ here have different meaning from those in the original eight-band model~(\ref{BdG}).
Under these symmetry constraints, the effective Hamiltonian takes the generic form of
\begin{equation}\label{effH}
H_\text{D}=k_x s_y\tau_x-k_y s_x\tau_x+k_z\tau_z\,.
\end{equation}

When the mirror symmetry is broken, the Dirac node loses its protection and thus may be gapped out. For example, a
mirror symmetry breaking perturbation $\delta s_z\tau_x$ in (\ref{effH}) leads to doubly degenerate gapped Dirac bands with dispersion
$\varepsilon=\pm\sqrt{k^2+\delta^2}$.
However, the surface Majorana Kramers pair is even robust against mirror symmetry breaking,
as long as the perturbation does not close the gap along the time-reversal and mirror invariant line with nontrivial $\gamma_m$.
This is because the parity of total $\gamma_m$ determines~\cite{zhang2013m} the $\mathbb{Z}_2$ index of a SC with TRS,
and in the current example odd $\gamma_m$ makes the $\mathbb{Z}_2$ index nontrivial.
Therefore, breaking the mirror symmetry may gap the Dirac nodes producing a topological SC~\cite{fu2010,sato} for which
the Majorana arc becomes the helical Majorana surface state, as seen in Fig.~\ref{fig2}(j-k).

When the inversion-gauge symmetry is broken, the nodal point is not required to be fourfold degenerate.
However, the spin Berry phases $\gamma_{\pm}=(\gamma_t\pm\gamma_m)/2$ are still well defined in mirror subspaces.
It is straightforward to obtain $\gamma_{\pm}=\pm1$ from the above analysis.
This implies there are two separate doubly-degenerate nodes in the mirror invariant plane,
one in each mirror subspace, as sketched in Fig.~\ref{fig2}(c).
In this plane, $\gamma_t$ and $\gamma_m$ are either the same or the opposite for any loop only including one doubly degenerate node.
Off this plane, even though $\gamma_m$ can no longer be defined, $\gamma_t$ is still well defined for any fully gapped loop since the chiral symmetry is unbroken.
Evidently, the two nodal points in the plane can extend off the plane and form a nodal ring,
which has to be normal to the plane as required by TRS and mirror symmetry, as sketched in Fig.~\ref{fig2}(c).
For example, such a symmetry breaking term $\delta \tau_x$ in (\ref{effH}) leads to a spectrum
$\varepsilon^2=\left(\sqrt{k_x^2+k_y^2}\pm\delta\right)^2+k_z^2$, which contains a nodal ring normal to the mirror invariant plane.
As a consequence on the surface,
the projected nodal ring has an extended area of zero energy modes inside~\cite{S1,S2}
and there is a surface arc connecting a pair of rings, as shown in Fig.~\ref{fig2}(f) and (h).
At a mirror symmetric surface, the surface arc must be dispersionless and spin degenerate.

A Dirac node may split into a pair of Weyl nodes~\cite{volovik,meng,sau,das,xu,go} only when TRS is broken.
In the presence of chiral symmetry, Chern number must be zero and hence an isolated Weyl node cannot be protected in a 3D SC .
Indeed, the product of TRS and PHS is a chiral symmetry.
Since PHS is intrinsic for any 3D SC, a Weyl node can only be protected when TRS is broken.
As an example, a symmetry breaking term $\delta s_x\tau_y$ in (\ref{effH}) splits the Dirac node into two Weyl nodes
with energy dispersion $\varepsilon=\pm\sqrt{(k_x\pm\delta)^2+k_y^2+k_z^2}$.
When the mirror symmetry is unbroken, no node can be protected in the mirror invariant plane due to the absence of chiral symmetry,
and thus the two Weyl nodes move off the plane normally in opposite directions.
Once the pair of Weyl nodes splits in momentum instead of in energy, the inversion-gauge symmetry pins them to zero energy.
These results, together with the symmetry breaking perturbations in the original Cu$_x$Bi$_2$Se$_3$ model are described in Fig.~\ref{fig2}.

All the above features are reminiscent of ABC-stacked graphene films and recently discovered 3D Dirac semimetals,
in which Dirac nodes are protected by sublattice symmetries and a set of crystalline constraints, respectively.
Analogically, the nodal phase described by Eqs.~(\ref{BdG}) and (\ref{effH}) should be entitled ``Dirac SCs in 3D''.
We also note that when $\gamma_m$ changes by $N$ from $k_y=0$ to $k_y=\pi$ in the mirror invariant plane,
a pair of nodal points with $k^N$ dispersion appears and in general each node can split into $N$ Dirac nodes.
Consequently, the degeneracy of surface Andreev arc between two neighboring projected nodes
decreases by two successively from the surface BZ center to the corner.

The presence of surface Andreev arc for a Dirac SC depends on the surface orientation.
For example, in Fig.~\ref{fig1}(f) we plot the energy spectrum with $(010)$ surface termination.
Because the two bulk nodes project to the same point in the surface BZ, the surface Andreev arc shrink to a point,
which is in sharp contrast to the case for $(001)$ surface.
For those surfaces breaking the mirror symmetry, the surface Andreev arc is not protected and may completely disappear.
As a result, surface density of states in tunneling experiment differs between different surfaces, as shown in Fig.~\ref{fig1}(g).
Thus, to identify a topological nodal phase with surface sensitive probes, it is necessary to measure multiple surfaces
by angular resolved photoemission spectroscopy, scanning tunneling microscope, or anomalous thermal (spin) Hall effect.

The 3D Dirac SC may be realized in the nodal phase of Cu$_x$Bi$_2$Se$_3$,
which serves as a concrete example to illustrate the essential physics in this Letter,
though the precise pairing symmetry of Cu$_x$Bi$_2$Se$_3$ is still under hot debate~\cite{fu2010,hor,wray,hao,krie1,das2011,krie2,sasa,kirz,bay,chen,laws,hsie,peng,levy,kond,zoch,yip}.
Recently, there is a hint of the existence of nodal points in the specific heat data of Cu$_x$Bi$_2$Se$_3$
at high $x$ values~\cite{private}. As suggested by our theory, different symmetry breaking in different samples may explain why different groups have observed different phases in Cu$_x$Bi$_2$Se$_3$.
The search for other candidate materials goes well beyond the scope here
and deserves a separate study in the near future.

{\indent{\em Acknowledgments.}}---
The authors are indebted to Y. Ando, D. L. Deng, L. Fu, C. L. Kane, E. J. Mele, A. Rappe, A. Schnyder, and S. Zaheer
for helpful inspirations and discussions. This work was supported by SUTD-SRG-EPD-2013062, NSFC Grant No. 11174022,
DARPA Grant No. SPAWAR N66001-11-1-4110 (UPenn), and NSF Grant No. NSF-PHY11-25915 (KITP).


\begin{references}
\bibitem{regular} Here as usual, for the calculation of Berry phase, a small regularization ($\propto k^2$) of the mass term is needed such that the mass term at $k=\infty$ recovers the trivial vacuum.
\bibitem{pi}            At $k_y=\pi$ the gap must be trivial and connected to the vacuum state.
\bibitem{parameter-fig1} The parameters used are $v=1.5$, $v_z=1$, $m=0.8$, $\mu=1$, and $\Delta=0.2$. In Fig.~\ref{fig2}(i), we take $m=0$ and $\mu=0.4$ for better visualization.
\end{references}
\end{document}